\documentclass[a4paper,fleqn]{cas-dc}

\usepackage[numbers]{natbib}
\usepackage{color}
\usepackage{graphicx}
\usepackage{epstopdf}
\usepackage{soul}

\colorlet{mygreen}{green!60!gray}

\newcommand{\QQ}{\mathcal Q}
\newcommand{\UU}{\mathcal U}

\def\tsc#1{\csdef{#1}{\textsc{\lowercase{#1}}\xspace}}
\tsc{WGM}
\tsc{QE}
\tsc{EP}
\tsc{PMS}
\tsc{BEC}
\tsc{DE}

\begin{document}
\let\WriteBookmarks\relax
\def\floatpagepagefraction{1}
\def\textpagefraction{.001}
\shorttitle{ST-DM watermarking of DNN}
\shortauthors{Yue Li et~al.}

\title [mode = title]{Spread-Transform Dither Modulation Watermarking of Deep Neural Network}




\author[1]{Yue Li}
\cormark[1]
\ead{liyue859000040@my.swjtu.edu.cn}
\credit{Research, paper writing, experiments}

\address[1]{School of Information Science \& Technology, Southwest Jiaotong University, 611756 Chengdu, China}

\author[2]{Benedetta Tondi}
\ead{benedettatondi@gmail.com}

\credit{Supervision of experiments, paper writing}

\author[2]{Mauro Barni}
\ead{barni@dii.unisi.it}

\credit{Coordination, Conceptualization}

\address[2]{Department of Information Engineering and Mathematics, University of Siena, Via Roma, 53100 Siena, Italy}



\cortext[cor1]{Corresponding author}


\begin{abstract}
DNN watermarking is receiving an increasing attention as a suitable mean to protect the Intellectual Property Rights associated to DNN models. Several methods proposed so far are inspired to the popular Spread Spectrum (SS) paradigm according to which the watermark bits are embedded into the projection of the weights of the DNN model onto a pseudorandom sequence. In this paper, we propose a new DNN watermarking algorithm that leverages on the watermarking with side information paradigm to decrease the obtrusiveness of the watermark and increase its payload. In particular, the new scheme exploits the main ideas of ST-DM (Spread Transform Dither Modulation) watermarking to improve the performance of a recently proposed algorithm based on conventional SS. The experiments we carried out by applying the proposed scheme to watermark different models, demonstrate its capability to provide a higher payload with a lower impact on network accuracy than a baseline method based on conventional SS, while retaining a satisfactory level of robustness.
\end{abstract}



\begin{keywords}
ST-DM \sep Watermarking \sep DNN
\end{keywords}

\maketitle




\section{Introduction}

Due to their outstanding performance, Deep Neural Networks (DNN) are increasingly used and commercialised in virtually all application scenarios wherein complex data for which precise statistical models do not exist must be analysed and processed. Training a DNN model is a difficult and computational intensive piece of work, requiring huge amounts of labelled data and extensive training procedures that may easily go on for weeks, even on powerful workstations equipped with several GPUs. For this reason, the demand for methods to protect the Intellectually Property Rights (IPR) associated to DNN is raising. Following similar efforts made for media protection \cite{BBCP98, Bloom99, Pod01}, the use of watermarking has recently been proposed as a way to track the distribution of DNN models and protect the IPR of model vendors \cite{AdiUsenix18,Uchida17}. By indissolubly tying a watermark to a DNN model, in fact, it would be possible to prove the ownership of the model, and to track its unauthorized distribution, with the possibility of tracing back to the user who illegally distributed it.

With respect to multimedia watermarking, for which a well-established theory has been developed in the last two decades \cite{Cox02, BB04}, embedding a watermark into a DNN model is quite a different piece of work, for which only a few newly proposed techniques exist \cite{Uchida17, darvish2019deepsigns, zhang2018protecting, le2019adversarial}. Generally speaking, DNN watermarking techniques can be split in two main categories: static and dynamic watermarking. Static DNN watermarking methods, like \cite{Uchida17} \cite{chen2019deepmarks}, embed the watermark into the weights of the DNN model. With dynamic watermarking, instead, the watermark is associated to the behaviour of the network in correspondence to specific inputs. For instance, the watermark may be associated to the activation map of the neurons in correspondence to certain inputs, as in \cite{darvish2019deepsigns}, or to the final output of the model, as in \cite{le2019adversarial, zhang2018protecting, chen2019blackmarks}.

With regard to static DNN watermarking, which is the kind of technique we focus on in this paper, a possible approach to embed the watermark consists in adding a specific term to the loss function used for training, requiring that the weights of the network satisfy certain properties. For instance, in \cite{Uchida17}, it is required that the weights {\em correlate well} with a pseudorandom watermark sequence. In this way, the watermark is not added to a pre-trained model, on the contrary the weights are trained from scratch in such a way to satisfy the desired properties. Directly generating the weights so that they have a high correlation with the watermark sequence somewhat resembles spread spectrum watermarking with informed embedding \cite{Mill04}, according to which embedding is achieved by applying a signal-dependent perturbation to the to-be-watermarked sequence.
In this way, the embedding procedure is adapted to the signal at hand resulting in a lower distortion and a lower (ideally zero) bit error rate.

Informed coding is another concept that has been proven to greatly improve the performance of a watermarking system \cite{Mill04}. The informed coding paradigm stems from the interpretation of watermarking as a problem of channel coding with side information at the transmitter \cite{Cox99, Costa83}. In a nutshell, with informed coding each watermark message is associated to a pool of codewords (rather than to a single one), then informed watermark embedding is applied by choosing the codeword that results in the minimum distortion. Practical implementations of the informed  coding paradigm include several popular watermarking schemes like Quantization Index Modulation (QIM) \cite{Chen01}, Dither Modulation (DM) \cite{Chen98} and Scalar Costa's Scheme (SCS) \cite{Egg03}. The possibility of coupling the strengths of QIM and Spread Spectrum (SS) watermarking has also been investigated leading to Spread-Transform Dither-Modulation (ST-DM) watermarking \cite{Chen01}, whereby DM is applied to the projection of the watermark sequence onto a predefined spreading direction. Within the wide class of informed watermarking methods, ST-DM has been widely applied due to its simplicity and good performance.

In this work, we propose a new DNN watermarking algorithm based on ST-DM. To do so, we modify the algorithm proposed in \cite{Uchida17} by training the DNN with a new loss function. The new loss is explicitly designed in such a way that the correlation of the DNN weights with a given spreading sequence assumes a quantized value belonging to one of two interleaved and equally spaced quantizers, associated, respectively, to a watermark message bit equal to 0 or 1. Multi-bit watermarking, is then achieved by using different spreading sequences to encode different bits. According to watermarking theory \cite{BB04}, the expected advantage of the informed coding paradigm is either a lower obtrusiveness of the watermark for a given payload, or the possibility of embedding a higher payload for a given level of obtrusiveness. On the other hand, the application of DM on top of spread transformation, that is after projection on a pseudorandom spreading sequence, guarantees that a satisfactory level robustness is achieved.

In order to verify the effectiveness of the proposed approach, we applied the new algorithm to watermark different DNN architectures targeting different classification tasks. By comparing the results we got with those achieved by Uchida et al's algorithm \cite{Uchida17}, the advantages predicted by theory are confirmed in terms of a larger achievable payload and a smaller impact on the accuracy of the network. We also assessed the robustness of the watermark, confirming that the robustness loss with respect to conventional SS watermarking is a minor one.

The rest of this paper is organised as follows. In Section \ref{sec.prior}, we review prior art in DNN watermarking. In Section \ref{sec.prop}, we briefly review ST-DM watermarking and describe our algorithm for ST-DM watermarking of DNN models. Sections \ref{sec.exp} and \ref{sec.res} are devoted to the experimental validation of the new algorithm. The paper ends in Section \ref{sec.con} with our conclusions and the discussion of some directions for future research.

\section{Prior art and Background}
\label{sec.prior}


Digital watermarking aims at embedding a
message into a digital content (an image, an audio file, etc \dots), in such a way that the message can be reliably recovered and used to demonstrate the ownership of the content, or trace back to the individual who redistributed the watermarked content illegaly.
In this section, we briefly review the existing literature on DNN watermarking, with particular attention to the methods proposed by Uchida et al's is \cite{Uchida17}, which is the basis of the new watermarking scheme proposed in this paper.

\subsection{Existing DNN watermarking techniques}

Several recent works have explored the possibility of injecting a watermark into a DNN model.
Watermarking of DNN models leverages the capability of DNNs to fit data with arbitrary labels. Such a capability is achieved thanks to a huge
number of parameters that can be also used to
carry additional information beyond what is required for the primary
classification task the network is dedicated to.

As we said in the Introduction, DNN watermarking techniques can be split into static and dynamic methods.

\paragraph{Static watermarking.} The first example of static DNN watermarking has been proposed by Uchida et al. in \cite{Uchida17, nagai2018digital},
according to which the watermarking bits are embedded into the
weights of the to-be-marked DNN.
Another static watermarking method has been proposed in \cite{chen2019deepmarks}. Both in \cite{Uchida17} and \cite{chen2019deepmarks}, embedding is achieved by adding a proper regularization term to the loss function during training. More details on the system described in \cite{Uchida17} are provided in the next section.
The maximum capacity that can be achieved with such watermarking schemes, without affecting the primary classification task, depends on the dimensionality and the number of parameters of the network.
Extracting the watermark requires
white-box access to the model in order to get the necessary information about the values of the model weights.
An obvious drawback with static methods is that the watermark, in general, is not very robust against model re-training or model pruning.

\paragraph{Dynamic watermarking.} With dynamic watermarking, the watermark bits are extracted by looking at the model outputs when the network is fed with a specific input, sometimes called triggering input. With these approaches,  the watermark can then be extracted in a black-box manner, since access to the internal status of the network is not required. The watermark may also be embedded in the activation map resulting from the application of the triggering input. In this way a higher payload can be embedded, however watermark extraction requires that the internal status of the network is accessible.
Dynamic watermarking methods  have been proposed in \cite{le2019adversarial,zhang2018protecting,AdiUsenix18,darvish2019deepsigns}.
They all focus on zero-bit watermarking, except for \cite{darvish2019deepsigns}, which proposes two dynamic watermarking methods, one for  zero-bit the other for multi-bit watermarking.
Specifically, the method proposed in \cite{le2019adversarial} consists in modifying the original
model boundary by using adversarial retraining; the watermark is given by the adversarial examples close to the decision boundary considered for the retraining. To extract the watermark, the model is queried by the adversarial  images (playing the role of a watermarking key). Zhang et al \cite{zhang2018protecting} propose to train the to-be-watermarked model with a set of input  crafted in order to trigger the assignment of a specific target label on those images. This approach is very similar to trojaning \cite{liutrojaning} and shares the same advantages and drawbacks. 
Similarly, Yossi et al.  \cite{AdiUsenix18} explore the possibility of using images that are misclassified by the model as the key images.
Finally, in \cite{darvish2019deepsigns}, Rouhani et al.  introduce a general watermarking methodology that
can be used in both white-box
and black-box settings, where watermark extraction may or may
not require to access the model internal parameters. The algorithm is based on embedding the watermark into the probability density function (pdf)
of the activations in various layers of the DNN.
The approach is shown to withstand various removal and
transformation attacks, including model compression, fine-tuning,
and watermark overwriting.

\subsection{Uchida et al's DNN watermarking algorithm}
\label{Uchida_paper}


Being the basis of the watermarking method proposed in this paper, we now describe in more details the watermarking algorithm introduced by Uchida et al. in \cite{Uchida17}.


Let us indicate with ${\bf b} \in \{0,1\}^l$, the vector with the watermark bits. For a selected convolutional layer, let $(s, s)$, $d$, and $n$ represent, respectively, the kernel size of the filters, the depth of the input and the number of filters. Ignoring the bias term, the weights of the selected layer can be denoted by a tensor $ \mathbf{W} \in \mathbb{R}^{s\times s\times d\times n}$.
In order to embed the watermark bits into the weights, it is convenient to flatten $\mathbf{W}$ according to the following steps: i)  calculate the mean of $\mathbf{W}$ over the $n$ filters, getting $\overline{\mathbf{W}} \in \mathbb{R}^{s\times s\times d}$ with $\overline{W}_{ijk} = \frac{1}{n} \sum_{h=1}^{n}W_{ijkh} $, in order to eliminate the effect of the order of the filters; ii) flatten $\overline{\mathbf{W}}$ producing a vector $\mathbf{w} \in \mathbb{R}^{v}$ where  $v=s\times s\times d$. Embedding the watermark bits into the weights corresponds to embed the vector $\mathbf{b}$ into the vector $\mathbf{w}$. Embedding is achieved by training the network with  a loss function $E(\mathbf{w})$ defined as follows:
\begin{equation}\label{loss_fuction1}
E(\mathbf{w}) = E_{0}(\mathbf{w}) + \lambda E_{R}(\mathbf{w}),
\end{equation}
where $E_{0}(\mathbf{w})$ represents the original loss function for the target DNN model (ensuring a good behavior with regard to the classification task),  $E_{R}(\mathbf{w})$ is a regularization term added to ensure correct watermark decoding, and $\lambda$ is a parameter adjusting the tradeoff between the original loss  term and the regularization term.  Specifically, $E_{R}(\mathbf{w})$ is given by
\begin{equation}\label{regulization}
E_{R}(\mathbf{w}) = -\sum_{j=1}^{l}(b_{j}\log(y_{j})+(1-b_{j})\log(1-y_{j})),
\end{equation}
\noindent where $b_{j}$ is $j$-th bit of $\mathbf{b}$ and $y_{j}=\sigma\big(\sum_{i}X_{ji}w_{i}\big)$, with $w_{i}$ denoting the $i$-th element of $\mathbf{w}$, $\mathbf{X} \in \mathbb{R}^{l\times v}$ playing the role of the watermarking key and
where $\sigma(\cdot)$ is the sigmoid function:
\begin{equation}
\sigma(x)=\frac{1}{1+\exp(-x)}.
\end{equation}
%

Three possible ways for the construction of $\mathbf{X}$ were considered in \cite{Uchida17}.
Eventually, based on experimental considerations, $\mathbf{X}$ is built by considering entries independently drawn from a standard normal distribution $N(0,1)$.
The algorithm to extract the watermark is pretty simple, since it consists in computing the projection of $\mathbf{w}$ onto each $X_j$,  and thresholding the projection at 0, that is:
\begin{equation}
\hat{b}_j =\begin{cases}
1 & \sum_{i}X_{ji}w_{i} \geq 0,\\
0 & \text{otherwise}.
\end{cases}
\end{equation}

\section{The proposed DNN watermarking algorithm}
\label{sec.prop}

\subsection{Security model and requirements}
\label{subsec.requirements}

We start by describing the security model and the requirements that the watermark message must  satisfy.

\paragraph{Security model.}

The watermarking method we aim at developing should be usable for both ownership verification and traitor tracing \cite{fiat2001dynamic}, to help verifying the legitimate owner of the model or tracing back to the individual who illegally redistributed it. These goals are more easily achieved by means of multi-bit watermarking, due to its superior flexibility with respect to zero-bit watermarking (see \cite{BB04} for a detailed discussion on the pros and cons of readable - or multibit - watermarking and detectable - or zero-bit - watermarking).

\begin{figure}[]
	\centering
	\includegraphics[scale=.35]{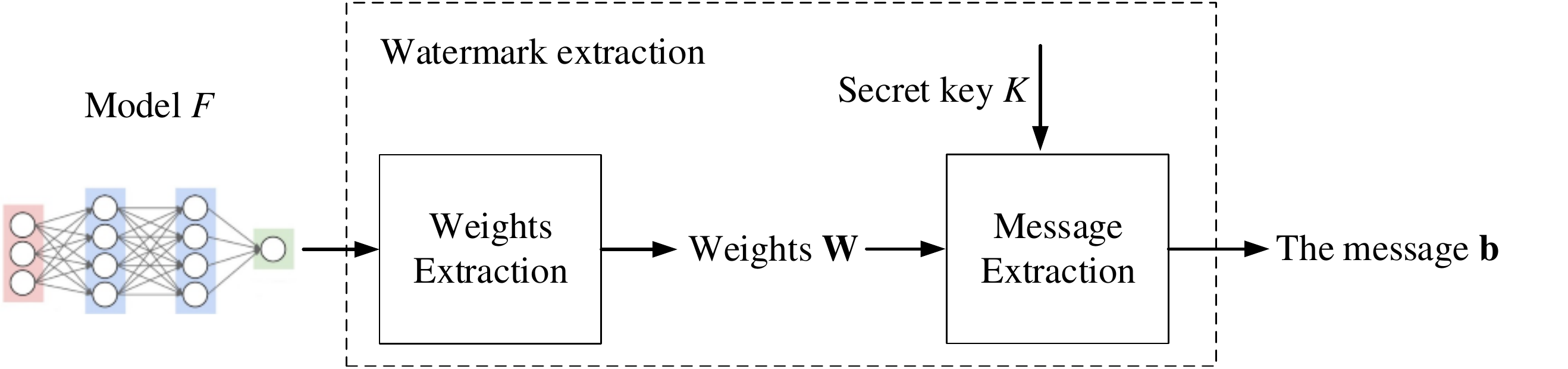}
	\caption{The watermark extraction process.}
	\label{watermark_extraction}
\end{figure}

Let $F$ be the to-be-protected model. During the training process, the model owner embeds a message $\mathbf{b}$ into the weights of $F$.
To do so, the embedder relies on a secret key $K$. The watermark extraction process is depicted in Figure \ref{watermark_extraction}.
As shown in the figure, watermark retrieval requires the knowledge of the key $K$ and is carried out by inspecting the weights of $F$, thus qualifying the proposed approach as a white-box watermarking algorithm. Watermark retrieval does not require that the message $\mathbf{b}$ is known in advance as it would have been in the case of zero-bit watermarking. As we said, the content of the watermark may be used to determine the owner of the model by means of an ownership verification protocol, or identify the individual who redistributed the content, by means of a traitor tracing protocol.

In the rest of the paper, we describe the insertion and extraction steps of the new ST-DM watermarking algorithm, and investigate the payload and robustness achieved by the new algorithm, without caring about the specific protocols wherein the watermark is used.

\paragraph{Watermark requirements} Based on the security model described above, it is necessary that the watermarking algorithm satisfies the following properties.
\begin{itemize}

\item{\textbf{\emph{unobtrusiveness}}: embedding the watermark message $\bf{b}$ should not have a significant impact on $F$. That is, the presence of the watermark should not degrade the accuracy of the network, with respect to a non-watermarked one.}
\item{\textbf{\emph{robustness}}: it should be possible to recover $\bf b$ also from a modified version of the network $F$. }
\item{\textbf{\emph{integrity}}: in the absence of model modifications, the extracted watermark $\bf b^{\prime}$ should be equal to $\bf b$, that is in the absence of modifications the bit error rate should be zero.}
\item{\textbf{\emph{payload}}: the payload of the watermark, that is the length of the message $\bf b$, should be as large as possible. This is a particular important requirement for traitor tracing applications, since a larger payload permits to index a larger number of users and allows the use of more powerful anti-collusion codes \cite{trappe2003anti}.
}
\end{itemize}

\paragraph{Robustness requirements.} With respect to the robustness requirement, we consider two kinds of model modifications: fine-tuning and parameter (network) pruning. These are common operations carried out routinely by network users, even in a non-adversarial setting wherein the users do not explicitly aim at removing the watermark from the model. 

\begin{itemize}

\item{\textbf{\emph{Fine-tuning}} is a common operation related to transfer learning. It consists in retraining a model that was initially trained to solve a given task, so to adapt it to solve a new task (possibly related to the original one). Computationally, fine-tuning is by far less expensive than training a model from scratch, hence it is often applied by model users to adapt a pre-trained model to their needs. Since fine-tuning alters the weights of the watermarked model, it is necessary to make sure that the watermark is robust against a moderate amount of fine-tuning.}

\item{\textbf{\emph{Network pruning}} is a common strategy to simplify a complicated DNN model to deploy it into low power or computationally weak devices like embedded systems or mobile devices. During pruning, the model weights whose absolute value is smaller than a threshold are cut-off to zero artificially. We require that the embedded watermark is resistant to this operation.}
\end{itemize}

\subsection{ST-DM watermarking}
\label{subsec.stdm}

In this section, we briefly review the main ideas behind ST-DM watermarking, since they represent the basis for the new DNN watermarking method proposed in this paper.

\begin{figure}[t!]
\centering
\includegraphics[scale=.40]{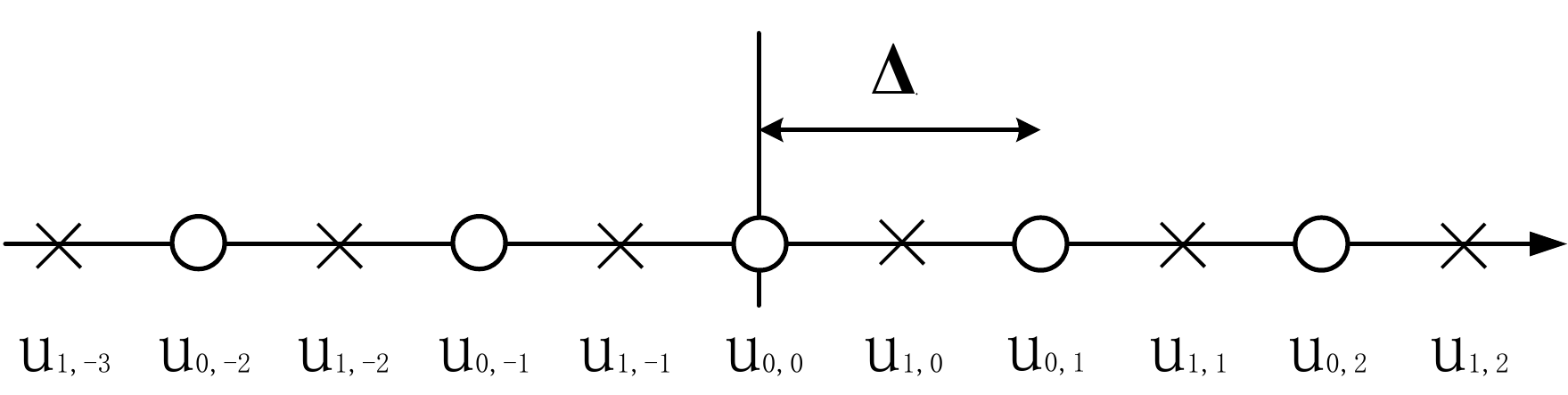}
\caption{Codebook entries for scalar DM watermarking.}
\label{fig.DM}
\end{figure}

Spread Transform Dither Modulation (ST-DM) is a watermarking algorithm coupling QIM and spread spectrum in a very simple fashion. The starting point for understanding ST-DM is Dither Modulation watermarking (DM), the simplest form of QIM. Given a host sample\footnote{In contrast to the common terminology adopted in watermarking literature, we use the symbol $w$ to indicate the samples hosting the watermark since here we are interested in watermarking the weights of CNN models.} $w$ and a watermark bit $b$, the marked sample $w_m$ is obtained by quantizing $w$ with one of two scalar quantizers $\QQ_0$ and $\QQ_1$. As shown in Figure \ref{fig.DM}, the codebooks associated to
$\QQ_0$ and $\QQ_1$ form two uniform interleaved quantizers with quantization step $\Delta$:
\begin{align}
    &\UU_0 = \left\{ k\Delta, k \in \mathbb{Z} \right\}\\
    &\UU_1 = \left\{ k\Delta + \Delta/2, k \in \mathbb{Z}
    \right\}.
    \label{eq.codebook_DM}
\end{align}

\begin{figure}
	\centering
	\includegraphics[scale=.43]{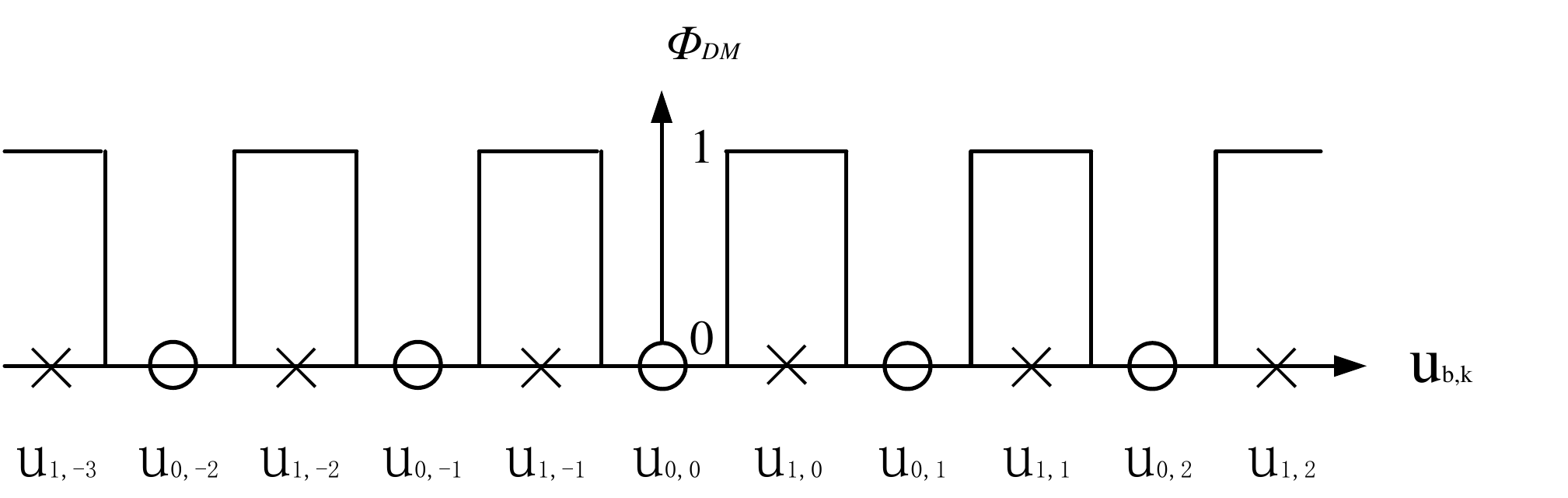}
	\caption{The decoding function $\phi _{DM}(w)$.}
	\label{fig.DMdec}
\end{figure}

Watermark embedding is achieved by quantizing $w$ either with $\QQ_0$ or $\QQ_1$:
\begin{equation}
    w_m = \left\{
    \begin{split}
        {\mathcal Q}_0(w)~~~~~&b=0\\
        {\mathcal Q}_1(w)~~~~~&b=1
    \end{split}
    \right.
    \label{eq.SDM}
\end{equation}
Retrieving the watermark from $w_m$ is straightforward. Given a watermarked sample $w_m$, it is only necessary to identify the entry in $\UU_0 \cup \UU_1$ closest to $w_m$ and see if such an entry belongs to $\UU_0$ or $\UU_1$, in formulas:
\begin{equation}
    \hat{b} = \phi_{DM}(w) = \arg \min_{b=0,1} (\min_{u_{k}\in {\mathcal U}_b} | w_m -
    u_{k}|).
    \label{eq.DM_decod_rule}
\end{equation}
A graphical representation of the decoding function
$\phi_{DM}(w)$ is given in Figure \ref{fig.DMdec}.

The intuition behind DM watermarking is that for every host sample $w$ a nearby codeword exists allowing to quantize $w$ with a small distortion. In this sense, $\Delta$ controls the distortion introduced by the watermarking process since a smaller $\Delta$ results in a smaller quantization step and hence a smaller distortion. A drawback with DM watermarking, especially when a small value of $\Delta$ is used,  is its lack of robustness. In fact, adding a small perturbation to $w_m$ may easily move the sample in the proximity of a wrong codebook entry thus producing a decoding error. ST-DM provides a way to increase the robustness of DM watermarking (at the price of a reduced payload) by embedding the bit $b$ into a sequence of $n$ host samples  ${\bf w} = (w_1, w_2 \dots w_n)$.  More specifically, let $\rho_x$ be the projection of ${\bf w}$ onto a unitary-norm pseudo-random sequence ${\bf x} = (x_1, x_2 \dots x_n)$:
\begin{equation}
    \rho_x = \left< {\bf w}, {\bf x} \right> = \sum_i w_i x_i.
\label{eq.STDM1}
\end{equation}
ST-DM watermarking works by applying DM to $\rho_x$ rather than to the samples in ${\bf w}$. More specifically, first the projection of ${\bf w}$ onto ${\bf x}$ is removed from ${\bf w}$, then a new component yielding the desired projection is added back to ${\bf w}$:
\begin{equation}
	{\bf w}_m = {\bf w} - \rho_x {\bf x} + \QQ_b(\rho_x) {\bf x}
\label{eq.STDM2}
\end{equation}

To embed more than one bit into ${\bf w}$, the above procedure is repeated for different pseudo-random directions:
\begin{equation}
	{\bf w}_m = {\bf w} - \sum_i \big( \rho_{x_i} {\bf x}_i + \QQ_b(\rho_{x_i}) {\bf x}_i \big).
\label{eq.STDM3}
\end{equation}
Watermark retrieval is obtained by computing the projections of ${\bf w}_m$ onto the pseudo-random directions ${\bf x}_i$ and applying DM decoding as in Eq. \eqref{eq.DM_decod_rule} to the projected values.
If the pseudo-random directions are orthogonal to each other, quantizations over different directions do not interfere with each other, hence resulting in error-free watermark retrieval. In practice, if the sequences are generated randomly and $n$ is big enough, they can be assumed to be nearly orthogonal and error-free decoding is still possible, unless the payload is too large.

\subsection{ST-DM-based watermarking of DNN models}
\label{subsec.embedding}



We are now ready to describe the new DNN watermarking algorithm. Our goal is to leverage on the superior performance allowed by ST-DM watermarking to find a better trade-off between watermark unobtrusiveness and payload. To do so, we modify the scheme proposed by Uchida et al. to incorporate within it the ST-DM watermarking principles.




By following the notation introduced in Section \ref{Uchida_paper},
we embed the vector ${\bf b}$ into $\mathbf {w}$, by training the DNN model so that applying the watermark decoding function to $\mathbf{w}$ results in the correct decoding bits. In particular, the loss function used for training
has the following form:
\begin{equation}\label{loss_function2}
E_{\text{ST-DM}}(\mathbf {w})=E_{0}(\mathbf{w})+\lambda E_{R}'(\mathbf{w})
\end{equation}
where $\lambda$ controls the tradeoff between
the original loss term and the new regularization term $E_{R}'$.
The term $E_{R}'$, enforcing correct decoding of the watermark bits,
is given by
\begin{equation}\label{ours_regularization}
E_{R}'(\mathbf{w}) = -\sum_{j=1}^{l}(b_{j}\log(z_{j})+(1-b_{j})\log(1-z_{j})),
\end{equation}
%
where $\mathbf{z}$ corresponds to the application of the DM decoding function $\phi_{DM}$ to the projection of $\mathbf w$ over the pseudorandom directions determined by the rows of a pseudorandom matrix $\mathbf{X}$ playing the role of the watermarking key, that is:
\begin{equation}
z_j = \phi_{DM}(\sum_i X_{ji} w_i).
\end{equation}
To generate $\mathbf{X}$, we followed the same appoach used in \cite{Uchida17}, drawing the elements of $\mathbf{X}$ independently from a unitary normal distribution $N(0,1)$.

Since $\phi_{DM}$ is a highly non-linear function and hence it does not satisfy the regularity properties needed to apply back-propagation, we approximated $\phi_{DM}$ with a smoother function $\theta()$, defined as:
%
\begin{equation}\label{ours_function}
\theta(x) = \frac{e^{\alpha \sin \beta x}}{1 + e^{\alpha \sin \beta x}}.
\end{equation}
Then, $z_j$ in  Eq. \eqref{ours_regularization} is defined as $z_j = \theta\left(\sum_i X_{ji} w_i\right)$.

\begin{figure}
	\centering
	\includegraphics[scale=.40]{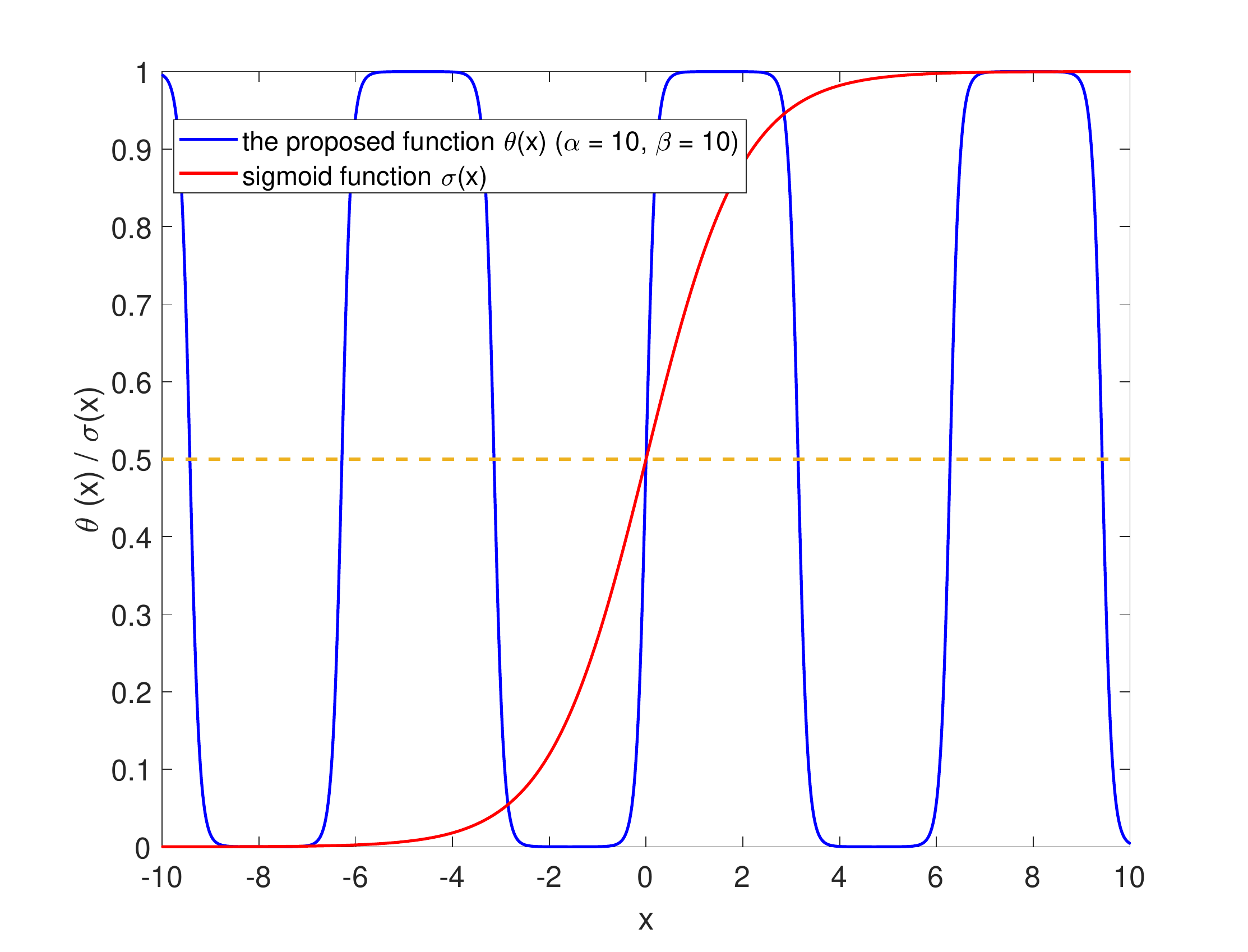}
	\caption{Behavior of the proposed function $\theta()$ and the sigmoid function used in \cite{Uchida17}.} 
	\label{two_function}
\end{figure}

The behavior of $\theta(x)$ is shown in Figure \ref{two_function}, for the setting $\alpha = 10$ and $\beta = 10$. In particular, $\beta$ controls the period of $\theta(x)$ and hence plays a role similar to $\Delta$ in standard ST-DM, while $\alpha$ controls the smoothness of the loss function, with large $\alpha$'s approximating better the rectangular shape of $\phi_{DM}$ at the expense of a lower smoothness.

The benefit of using the new loss function based on $\theta$ instead of the usual sigmoid function  as in \cite{Uchida17} can be easily understood by inspecting Figure \ref{two_function}. To binarize the outputs of $\theta()$ and $\sigma()$, we use the threshold value 0.5, then the bit is decoded as 0 when the output is lower than 0.5, as 1 otherwise.
Therefore, it is immediate to argue that with the new method the weights $w_i$  needs less modification on the average to achieve the same embedding target bit, with respect to \cite{Uchida17}.
As an example, if the initial values of $\sum_{i}X_{ji}w_{i}$ are equal to $-8$, corresponding to $\sigma\left(\sum_{i}X_{ji}w_{i}\right) = \theta\left(\sum_i X_{ji} w_i\right) = 0$ (see Figure \ref{two_function}), and we want to embed a bit '1', with the function $\theta()$ it is sufficient to reach the condition $\sum_i X_{ji} w_i^\prime = -6$, while with the sigmoid function the condition $\sum_{i}X_{ji}w_{i}^\prime \geq 0$ has to be reached for a successful embedding. 



\subsection{Extraction}
\label{subsec.extract}

To extract the watermark, we simply project the watermarked weights $\mathbf{w}_m$ 
onto the direction established by the projection matrix $\mathbf{X}$, apply the function $\theta()$, and then binarize the output by  thresholding at 0.5. Formally,
the $j$-th bit is extracted as:
\begin{equation}
\hat{b}_j =\begin{cases}
1 & \theta\left(\sum_{i}X_{ji}w_{i}\right) \geq 0.5,\\
0 & \text{otherwise}.
\end{cases}
\end{equation}

\section{Experimental setting}
\label{sec.exp}
Before discussing the performances of ST-DM watermarking, we discuss the setting used in our experiments. Such setting has been defined so to allow a fair comparison with SS watermarking, as implemented in \cite{Uchida17}. 




\subsection{Datasets and host networks} In our experiments, we evaluated the performance of the proposed watermarking method
mainly by referring to the CIFAR-10~\cite{krizhevsky2009learning} classification task, solved by using the popular Wide Residual Networks (WRN) model  \cite{zagoruyko2016wide}. To prove the generality of the proposed method, we also carried out some tests by considering another network architecture for the same task, that is ResNet~\cite{he2016deep}, and other two tasks, namely, the German Traffic Sign Recognition Benchmark (GTSRB) \cite{stallkamp2012man} and Imagenet~\cite{deng2009imagenet} tasks. For these two tasks, we considered the ResNet~\cite{he2016deep} and VGG~\cite{simonyan2014very} networks respectively.

\subsection{Settings and experiments}  WRN is an efficient variant of the residual network (ResNet) with decreased depth and increased width.
The structure of WRN is described in Table \ref{WRNs}.
Groups of convolutions are shown in brackets where $N$ is the number of blocks in each group.
The network width is determined by a factor $k$,  that establishes the growth rate of the number of filters $n$ from one layer to the other (and then of the depth of the input).
Only the second layer of each convolutional block in \cite{zagoruyko2016wide} is actually considered for embedding (reported in bold in the table), as done in \cite{Uchida17}.
Then, the number of filters is $n =  16\times k$ for each layer in the conv 2 block, $n = 32 \times k$ for each layer in the conv 3 block, and  $n = 64 \times k$ for the conv3. The input depth $d$ has the same value of $n$ when the second layer
of each convolutional block is considered. The kernel size of the filters is fixed to $(s,s) = 3\times 3$ for every layer.
The number of embedding weights $v = s\times s \times d$ in each layer of convolutional block is also reported in Table \ref{WRNs} (last column) as a function of $k$.
Following  \cite{zagoruyko2016wide}, for all the experiments we set $N=1$ and  $k=4$.
For simplicity, in the following,  we will refer to conv 2, conv 3, or conv 4 to denote the embedding layer, keeping in mind that only the second layer of each convolutional block is actually considered.
The maximum number of parameters that can be used for embedding in conv 2, conv 3 and  conv 4  is then 576, 1152, and 2304, respectively.


For training the non-watermarked and watermarked WRN models,
we used SGD 
with Nesterov momentum equal to 0.9 and cross-entropy loss $E_0$, with minibatch size 64. A total number of 200 epochs was considered. The learning rate was initially set to  0.01, and then dropped by a factor of 0.2 at 60, 120 and 160 epochs.


In addition to embedding the watermark within a single layer, as done in  \citep{Uchida17}, we also considered the case of multi-layer embedding, so to increase the number of weights hosting the watermark and allow a higher payload.

Finally, the  trade-off parameter $\lambda$'s in Eq.~(\ref{loss_fuction1}) and Eq.~(\ref{loss_function2})  are both set to 0.01. We refer to Section \ref{sec.choice-parameters} for the choice of the parameters $\alpha$ and $\beta$ of the ST-DM watermarking method.

To demonstrate the effectiveness on different architectures, we applied the proposed method (as well the method in \cite{Uchida17}) on ResNet and VGG. By following the instruction in \cite{he2016deep}, we considered ResNet-34 and ResNet-50. For ResNet-34, we considered 33 convolutional layers, and 49 convolutional layers for ResNet-50. For training, the learning rate was initially set to $10^{-3}$ for these two architectures, and then decreased by a factor 0.1 at 40, 80, 100 epochs, for a total number of 120 epochs and a minibatch size of 32. Regarding the trainable weights to be used for watermark embedding, we chose the weights of the second to last convolutional layer of ResNet-34 and ResNet-50, for which $v = 576$.

With regard to VGG, we adopted VGG-16 with 13 convolutional layers according to the architecture described \cite{simonyan2014very}. The learning rate was initially set to $10^{-2}$, and then decreased by a factor of 0.1 when the validation accuracy stops improving. The maximum number of epochs was set to 100. Besides, we used SGD with Nesterov momentum equal to 0.9 and cross-entropy loss $E_0$, with a batch size set to 256. For watermarking, we chose the second convolutional layer in block3, for which the number of embeddable weights $v$ is 2304.

\begin{table}[width=1\linewidth,cols=4,pos=h]
	\caption{Structure of WRN. The layers considered for embedding are highlighted in bold.}
	\label{WRNs}
	\begin{tabular*}{\tblwidth}{@{} CCCC@{} }
		\toprule
		Group  & Output size & Building block & $v$\\
		\midrule
		conv 1 & $32 \times 32$ & [$3 \times 3, 16$] & N/A \\
		conv 2 & $32 \times 32$ & $\begin{bmatrix}3 \times 3, 16 \times k\\ {\bf 3 \times 3, 16 \times k}\end{bmatrix} \times N$ & $\begin{matrix} 144 \times k \\ {\bf 144 \times k} \end{matrix} $ \\
		conv 3 & $16 \times 16$ & $\begin{bmatrix}3 \times 3, 32 \times k\\ {\bf 3 \times 3, 32 \times k} \end{bmatrix} \times N$ &  $\begin{matrix}  144 \times k\\  {\bf 288 \times k} \end{matrix} $  \\
		conv 4 & $8 \times 8$ & $\begin{bmatrix}3 \times 3, 64 \times k\\ {\bf 3 \times 3, 64 \times k}\end{bmatrix} \times N$ & $\begin{matrix}  288 \times k \\ {\bf 576 \times k} \end{matrix} $  \\
		 & $1 \times 1$ & avg-pool, fc, soft-max & N/A \\
		\bottomrule
	\end{tabular*}
\end{table}

\subsection {Choice of the parameters} 
\label{sec.choice-parameters}
Before showing and discussing the results we got, we pause to discuss the choice of the parameters defining the the loss term $E_R'$ of our method and  how we generalized $E_R$ in \citep{Uchida17} for a more fair comparison.

\begin{figure}
	\centering
	\includegraphics[scale=.40]{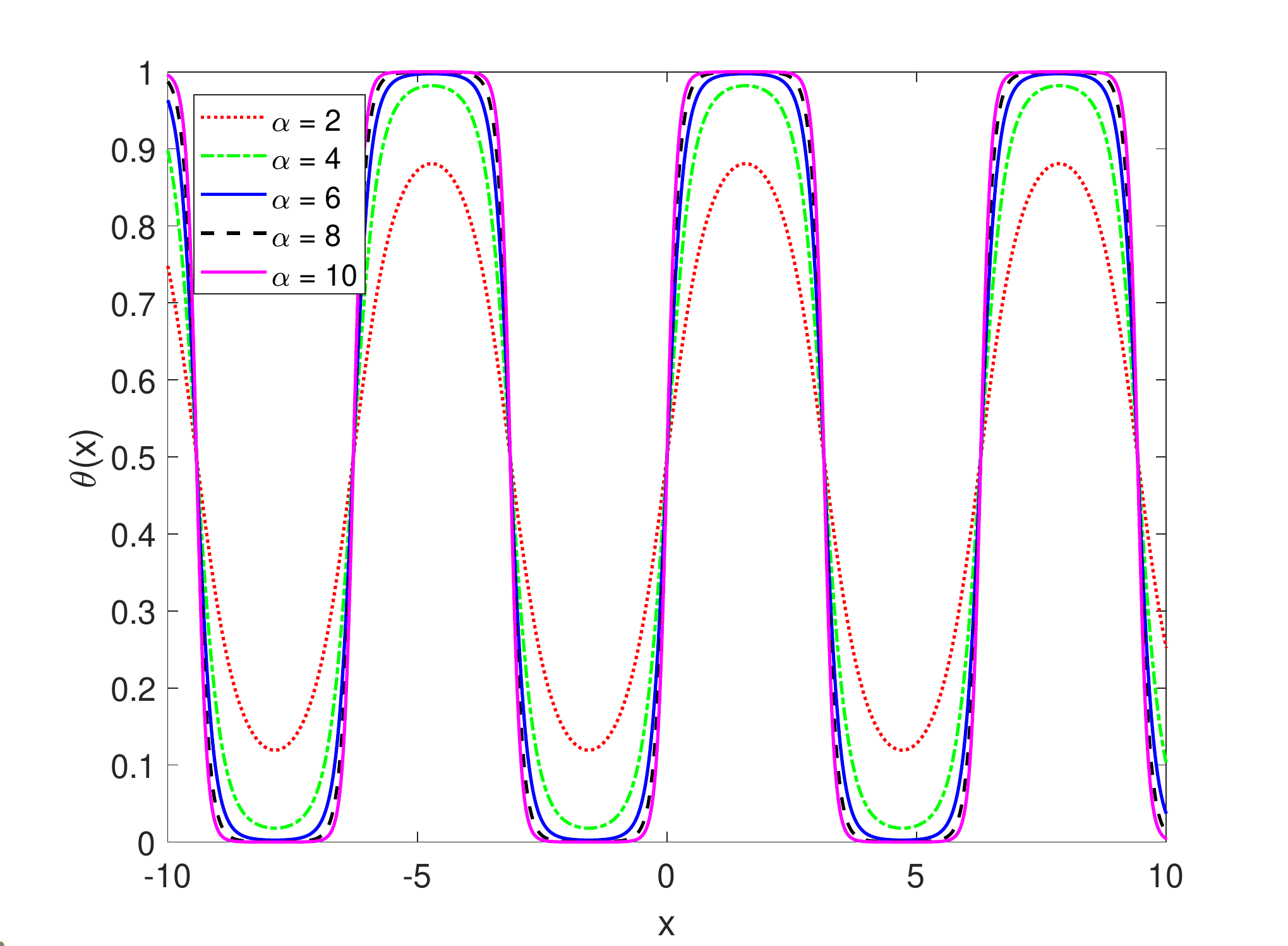}
	\caption{Activation function $\theta(x)$ with fix $\beta$ and varying $\alpha$}\label{simulation of theta}
\end{figure}

As discussed in Section \ref{subsec.embedding}, the choice of the parameters $\alpha$ and $\beta$ determining the shape of the function $\theta()$  in Eq. (\ref{ours_function}) is an important one.
A general view of the impact of $\alpha$ on  $\theta(x)$ is illustrated in Figure \ref{simulation of theta} for a fixed $\beta$ ($\beta = 1$).
We see that $\alpha$ has a significant influence on controlling the smoothness of the loss function.
For larger $\alpha$, $\theta(x)$ approximates better the rectangular shape of $\phi_{DM}$. However, too large values of $\alpha$
may result in an eccessive sensitivity of the loss to small variations of the weights, leading to training instability problems.
For our experiments, we tested several values of $\alpha$  and empirically set it to 10, which represents a reasonably large value (approximating well $\phi_{DM}$) that still guarantees the convergence of the models.

With $\alpha$ fixed to $10$, we optimized the value of $\beta$ by training with watermark embedding
and measuring the Test Error Rate (TER) of the watermarked model and the Bit Error Rate (BER) of the extracted watermark.
Table \ref{activation function with beta} reports the results
of the WRNs model watermarked with the proposed ST-DM embedding scheme (W-DM-WRN) for the CIFAR-10 task. The watermark was embedded in the Conv 2 layer with a  payload of 1024 bits.
We argue that the best $\beta$ leading to a small test error and to a zero BER is 10 (we found that this value is good also in other settings).
Then, in all our experiments we let $\alpha =\beta = 10$.

\begin{table}[width=.9\linewidth,cols=4,pos=h]
	\caption{TER  of W-DM-WRN and BER of the extracted watermark for different $\beta$ values ($\alpha = 10$).}\label{activation function with beta}
	\begin{tabular*}{\tblwidth}{@{} CCC@{} }
		\toprule
		 $\beta$ & TER(\%) & BER(\%)\\
		\midrule
		1 & 8.68 &4.69\\
		5 & 8.10 & 3\\
		10 & 7.63 & 0\\
		15 & 9.68 & 0\\
		20 & 9.58 & 0\\
		30 & 10 & 48.44 \\
		50 & 9.93 & 50.98\\
		
		\bottomrule
	\end{tabular*}
\end{table}

\begin{figure}
	\centering
	\includegraphics[scale=.50]{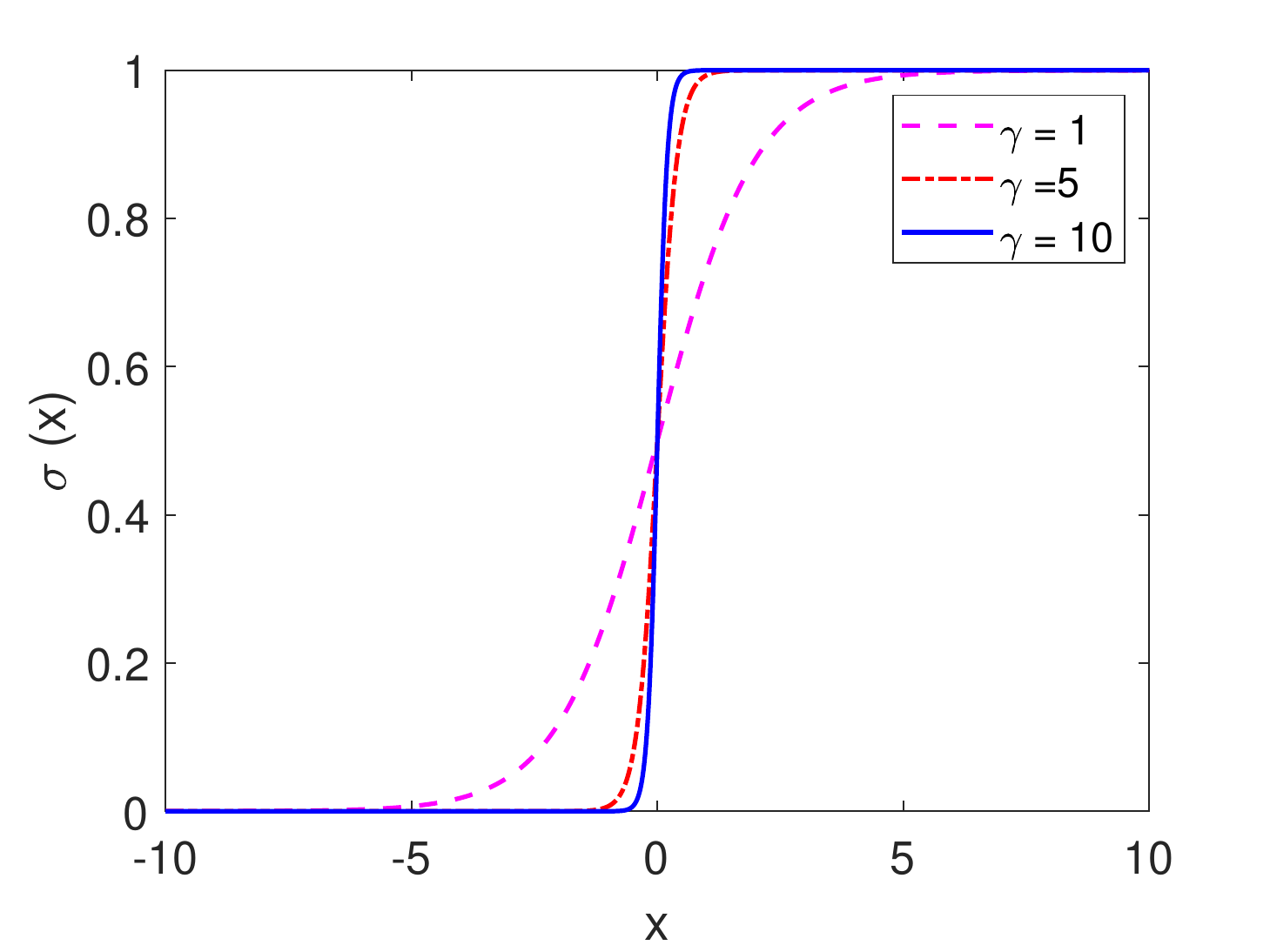}
	\caption{The sigmoid function in Eq. \eqref{eq.sigm_gen} for various $\gamma$.}
	\label{transformed_sigmoid}
\end{figure}

Since we optimize the parameters for the decoding function of the  proposed method, for a more fair comparison with the watermarking scheme in \citep{Uchida17}, we considered a generalization of the sigmoid function used therein and optimized it in a way similar to what we did for $\alpha$ and $\beta$. Specifically, the generalized sigmoid function is defined as:
\begin{equation}
\label{eq.sigm_gen}
\sigma(x) = \frac{1}{1 + e^{- \gamma x}}
\end{equation}
where $\gamma$ is a tunable parameter, determining the slope of the sigmoid function, in a way similar to $\alpha$ for $\theta$ (see Figure \ref{transformed_sigmoid}).
The performance of the watermarked WRN (W-WRNs) for several values of $\gamma$,  for the same choice of the embedding layer and payload we used for W-DM-WRN, are reported in Table \ref{gamma_experiments}. As it can be seen, the case $\gamma = 1$ adopted in \citep{Uchida17} is not the best one, and a much lower test error rate and BER  can be achieved by considered a larger $\gamma$.
Since BER = 0 for all $\gamma \ge 10$, while the test error does not change much by increasing $\gamma$ above 10, in our experiments we let $\gamma = 10$.

\begin{table}[width=.9\linewidth,cols=4,pos=h]
	\caption{TER of W-WRN and BER of extracted watermark  with different $\gamma$ values}
	\label{gamma_experiments}
	\begin{tabular*}{\tblwidth}{@{} CCC@{} }
		\toprule
		$\gamma$ & TER (\%) & BER(\%)\\
		\midrule
		1 & 12.11 &7.32\\
		5 & 10.5 & 0.49\\
		10 & 8.90 & 0\\
		15 & 8.92 & 0\\
		20 & 8.95 & 0\\
		30 & 8.88 & 0 \\
		50 & 8.99 & 0\\
		
		\bottomrule
	\end{tabular*}
\end{table}

\section{Analysis of the experimental results}
\label{sec.res}

According to watermarking theory,  ST-DM watermarking is expected to provide superior performance with respect to SS watermarking in terms of payload and unobtrusiveness, at the price of a possible loss of robustness. The goal of this section is to report the results of the experiments we carried out to demonstrate the gain of our new algorithm with respect to prior art, from the point of view of increased payload and reduced unobtrusiveness. We also report some experiments showing that, with a proper setting of the watermark parameters, such a gain can be achieved with no significant loss of robustness.

All the networks were trained from scratch both with and without watermark embedding, to evaluate the drop in the classification accuracy due to the presence of the watermark.
Specifically, the performance of the watermarked models are evaluated by measuring the test error rate of the model
for the envisaged task (thus assessing the unobtrusiveness of the watermark), where Test Error Rate (TER) = 1 - Accuracy, and the Bit Error Rate (BER) pertains to watermark extraction (thus assessing the watermark accuracy).

\subsection {Payload, bit error rate and test error rate}


We carried out our experiments for the CIFAR-10 task by considering various payloads and embedding the watermark into several convolution layers.
The test error rate of the baseline non-watermarked WRN model is 8.73\%.
The results we got are reported in Table \ref{payloads_experiments}(a)-(d) for different payloads and different host convolutional layers (conv 2 (a), conv 3 (b), conv 4 (c) and multi-layer embedding (d)).
In the case of multi-layer embedding,  the watermark is embedded into conv 2, conv 3 and conv 4 simultaneously. In this case, the  payload  of 7168 bits is achieved by embedding 1024 bits, 2048 bits and 4096 bits in conv 2, conv 3 and conv 4 respectively, while the 8400 bits payload is obtained by embedding 1200, 2400 and 4800 bits in conv 2, conv 3 and conv 4.
For each case, the performance of the ST-DM watermarked model (W-DM-WRNs) and a model watermarked as in \cite{Uchida17} (W-WRNs) are reported.

By inspecting the tables, we see that both methods do not degrade the accuracy of the original non-watermarked model, the test error rate  being even lower than the baseline.
This is possibly because the $L_2$ regularization terms  included for watermarking (see Eq. \eqref{loss_fuction1} and Eq. \eqref{loss_function2}) reduce overfitting, thus improving the test accuracy.
%
%
For large payloads, ST-DM watermarking has a significant advantage in the test error rate compared to Uchida et al's algorithm, see, for instance, the case of 1024 bits embedded in conv 2, 2048 bits embedded in conv 3 and 4096 bits embedded  in conv 4, where the gain is larger than 1\%.
More importantly, we observe that ST-DM watermarking works well, yielding BER = 0, also with larger payloads, i.e., when the payload is larger than the available number of training parameters for the selected layer (see last row of Tables \ref{payloads_experiments}(a), (b) and (c)).
For instance, with the new ST-DM algorithm we can embed 2400 bits in conv 3 with BER = 0 and a low test error while with W-WRN the BER is above 12\%. Not surprisingly, the gain of the  W-DM-WRN over W-WRN is even more evident in the case of multi-layer embedding.

\begin{table}[width=1.0\linewidth,cols=4,pos=h]
	\caption{TER and BER of W-DM-WRN (our) and W-WRN (Uchida et al, \cite{Uchida17})
for various payloads, considering different embedding layers.}
	\label{payloads_experiments}
	(a) Watermark embedding in Conv 2 \\
	(trainable parameters available for watermarking: 576 bits)
\begin{tabular}{c|c c | c c}
\hline
		Payload &  \multicolumn{2}{c|}{W-DM-WRN} &  \multicolumn{2}{c}{W-WRN}  \\ \cline{2-5}
		(bit)& TER(\%) & BER(\%) & TER(\%) & BER(\%)\\
\hline
		256 & 8.20 & 0 & 8.15 & 0\\
		512 & 7.75 & 0 & 8.28 & 0\\
		1024& 7.63 & 0 & 8.90 & 0\\
		1200& 7.96 & 0 & 8.05 & 15.67 \\
\hline
	\end{tabular} \\[12pt]

	(b) Watermark embedding in Conv 3\\
	(trainable parameters available for watermarking: 1152 bits)

\begin{tabular}{c|c c | c c}
\hline

		Payload &  \multicolumn{2}{c|}{W-DM-WRN} &  \multicolumn{2}{c}{W-WRN}  \\ \cline{2-5}
		(bit)& TER(\%) & BER(\%) & TER(\%) & BER(\%)\\
\hline
		256 & 8.15 & 0 & 8.02 & 0 \\
		512 & 8.07 & 0 & 8.30 & 0 \\
		1024& 7.79 & 0 & 8.41 & 0 \\
		2048& 7.85 & 0 & 8.93 & 0 \\
		2400& 8.22 & 0 & 8.22 & 12.46\\
\hline
	\end{tabular} \\[12pt]

	(c) Watermark embedding in Conv 4\\
	(trainable parameters available for watermarking: 2304 bits)
\begin{tabular}{c|c c | c c}
\hline
		Payload &  \multicolumn{2}{c|}{W-DM-WRN} &  \multicolumn{2}{c}{W-WRN}  \\ \cline{2-5}
		(bit)& TER(\%) & BER(\%) & TER(\%) & BER(\%)\\
\hline
		256 & 8.20 & 0 & 8.45 & 0 \\
		512 & 8.57 & 0 & 8.30 & 0 \\
		1024& 7.99 & 0 & 8.39 & 0 \\
		2048& 8.03 & 0 & 8.12 & 0 \\
		4096& 7.64 & 0 & 8.60 & 0\\
		4800& 8.65 & 0 & 8.25 & 11.88\\
\hline
	\end{tabular} \\[12pt]

	(d) Watermark embedding in multiple layers
\begin{tabular}{c|c c | c c}
\hline

		Payload &  \multicolumn{2}{c|}{W-DM-WRN} &  \multicolumn{2}{c}{W-WRN}  \\ \cline{2-5}
		(bit)& TER(\%) & BER(\%) & TER(\%) & BER(\%)\\
\hline
		7168 & 8.31 & 0 & 9.97 & 0 \\
		8400 & 8.19 & 0 & 9.75 & 13.10 \\
\hline
	\end{tabular}
\end{table}

We verified  that the advantage of the proposed scheme is confirmed when other network architectures and different tasks are considered.
Table \ref{payload_for_other_model} reports the test error rate and BER when we watermarked ResNet50, ResNet34 and VGG networks, trained on CIFAR-10, GTSRB and ImageNet respectively.
We see that the proposed scheme maintains its gain with respect to \cite{Uchida17} in terms of test error rate, while allowing to embed a larger payload with zero BER.
The advantage is expectedly larger in the multi-layer embedding case.


\begin{table*}[width=2.08\linewidth,cols=4,pos=h]
	\caption{Performance achieved with different DNN models and tasks. }\label{payload_for_other_model}
\begin{tabular}{c|c|c|c|c|cc|cc}
\hline
		Model & Dataset &  Embedding layer (number of & Baseline & Payload & \multicolumn{2}{c|}{W-DM-WRN} & \multicolumn{2}{c}{W-WRN} \\ \cline{6-9}
		& &  trainable parameters) & TER (\%) & (bits) &TER(\%) & BER(\%) & TER(\%) & BER(\%)\\
\hline
		\multirow {2}{*}{ResNet50}& \multirow{2}{*}{CIFAR-10} & \multirow {2}{*}{Penultimate conv layer (576)}& \multirow{2}{*}{7.51} & 512 & 7.08& 0 & 7.27 &0\\
		& & & &1000 & 7.49 &0 & 7.63& 12.85\\
		
\hline
		\multirow {2}{*}{ResNet34} & \multirow {2}{*}{GTSRB} & \multirow {2}{*}{Penultimate conv layer (576)} & \multirow {2}{*}{1.33} & 512 &1.19& 0& 1.49 &0\\
		& & & & 1024&0.96& 0 & 1.01 &13.09\\
\hline
		VGG16 & ImageNet & Block3 of conv2 (2304) & 8.75 &4096  & 7.83 &0 & 8.72 &0 \\
		\bottomrule
	\end{tabular}
\end{table*}

\subsection{Robustness evaluation}

In this section, we evaluate the robustness of the new proposed method against two types of very common unintentional attacks: fine-tuning and parameters pruning. Given that ST-DM watermarking is expected to be beneficial from a payload point of view with the risk of reducing the robustness of the watermark, the goal here is to show that the performance improvement described in the previous section is obtained with a negligible loss of robustness. As before, we use the scheme in \cite{Uchida17} as a baseline for our analysis.

\subsubsection {Robustness against fine-tuning}

Fine-tuning is the most common type of unintentional attack watermark DNN models are subject to, given its frequent use to adapt a pre-trained network to a new task. Applying fine-tuning to a pre-trained network, in fact, requires much less effort than training a network from scratch, and also avoids over-fitting when sufficient training data is not available for the new training.

To measure the robustness of the watermarked models against fine-tuning,
we considered the watermarked W-WRN and W-DM-WRN  models and fine-tuned them by re-training them with the standard loss (i.e, the cross-entropy loss $E_0$) for some more epochs.
%
The  parameters setting for fine tuning is left unchanged, except for the number of epochs that is set to 20. 
Table \ref{fine-tuning} shows the  results  we obtain by considering several groups of embedding layers and different payloads. In all cases we fine-tuned the models on the same CIFAR-10  dataset.
Not surprisingly, the test error slightly decreases after fine-tuning with the standard loss, since the weights are modified in such a way to increase the accuracy on the classification task; however, the watermark BER remains zero, so the fine-tuned models perform correct watermark decoding. We also verified that even if the fine tuning goes on for 120 epochs the BERs remain zero.


For further validation, we also fine-tuned watermarked W-DM-WRNs and W-WRNs models on a modified training set for the same CIFAR-10 task, that is by using different portions of the dataset for training and testing set.
For this experiments we considered the case of 256 bits embedded in conv4. The loss $E_0$ is considered for fine-tuning, which is carried out for 20 epochs. The results are shown in Table \ref{Finetuning with different datasets}.
We see that  both  watermarked models can resist to fine-tuning for 20 epochs, still achieving BER = 0. As an additional experiment, we also fine tuned the model on a different dataset, namely the GTSRB dataset. The results we got are reported in the same table. Not surprisingly, in this case the BER increases significantly for both models. Even if this is not a desired behaviour, W-DM-WRNs shares this weakness with W-WRNs, hence showing that this particular lack of robustness is not due to ST-DM. 


\begin{table*}[width=2.08\linewidth,cols=4,pos=h]
	\caption{Robustness against fine-tuning.} \label{fine-tuning}
	\begin{tabular}{c|c|ccc|ccc}
\hline
		Embedded & Payload & \multicolumn{3}{c|}{W-DM-WRN}  & \multicolumn{3}{c}{W-WRN} \\
\cline{3-8}
		layer & (bit) &  \multirow{2}{*}{TER (\%)}& TER & BER& \multirow{2}{*}{TER (\%)}& TER & BER\\
 &  &  & after attack(\%)&  after attack (\%) & & after attack (\%) & after attack (\%)\\
\cline{3-8}
		\hline
		Conv 2 & 256 & 8.20 & 8.02 & 0 & 8.15 & 8.11 & 0\\
\hline
		\multirow{2}{*}{Conv 3}& 256 & 8.15 & 8.05 & 0 & 8.02 & 7.58 & 0\\
		& 1024 & 7.79 & 7.65 &0&8.41& 8.17&0 \\
\hline
		\multirow{2}{*}{Conv 4}& 256 & 8.20 & 7.93 & 0 & 8.45 & 7.90 & 0\\
		& 4096 & 7.64 & 7.43 & 0 & 8.60 & 8.26 & 0\\
\hline
		multi-layer & 768 & 8.38 & 8.25 & 0 & 8.24 & 8.14 & 0\\
\hline
	\end{tabular}
\end{table*}

\begin{table*}[width=2\linewidth,cols=4,pos=h]
	\caption{Robustness against fine-tuning with different dataset.}\label{Finetuning with different datasets}
	\begin{tabular}{c|c c c| c c c}
		\hline
		\multirow{3}{*}{Dataset} &  \multicolumn{3}{c|}{W-DM-WRNs} &  \multicolumn{3}{c}{W-WRNs}  \\ \cline{2-7}
		& \multirow{2}{*}{TER (\%)} & TER & BER & \multirow{2}{*}{TER (\%)} & TER & BER(\%)\\
		& &after attack(\%) & after attack(\%) & & after attack(\%) & after attack(\%)\\
		
		\hline
		Modified CIFAR-10 & \multirow{2}{*}{8.20} & 6.83 & 0 & \multirow{2}{*}{8.45}& 6.27 & 0\\
		GTSRB& & 6.45 & 50.78 & & 8.59 & 41.80\\
		\hline
	\end{tabular}
\end{table*}

\subsubsection {Robustness against parameter pruning}

Considering the deployment of DNNs on mobile devices and other platforms with limited storage capability, parameter pruning is another unintentional attack that occurs in practical application.

In order to asses the robustness against parameter pruning, we randomly cropped a percentage $p\%$ of the $s\times s\times d\times n$ trainable parameters of the embedding layer for both W-WRN and W-DM-WRN models, by setting them to zero. Then, watermark extraction is carried out as usual.
%
The performance are assessed for several groups of embedding layers and pruning percentages $p$.

To show the influence of pruning on the TER, in Table \ref{Different pruning percentage} we report the results regarding W-WRN and W-DM-WRN with embedding layer  equal to conv 4 and a payload of 256 bits for different values of the pruning percentage. As shown in the table,
when the pruning percentage is equal to 30\%, the TER of the two models is already much higher than for the baseline model (8.73\%), however, both watermarking algorithms can resist even larger pruning percentages with a BER=0. From the table, we also see that W-DM-WRN starts having a non-zero BER when $p = 70\%$, while the BER of the W-WRN model is still 0, confirming the intuition that ST-DM watermarking is less robust than SS watermarking. However, the difference between the two systems is appreciable only for very large pruning percentage when the TER is unacceptably large. Hence, by any practical means, we can conclude that the robustness of W-WRN and W-DM-WRN against parameter pruning is the same.

%

The results we got for different embedding layers and payloads are reported in  Table \ref{pruning}.  We see that W-WRN and W-DM-WRN has a similar level of robustness against parameter pruning, and also the test error rate is similar.
For each case,
we considered the maximum pruning percentage $p$ before the test error increases too much (above the baseline), which is approximately the same for both W-WRN and W-DM-WRN models.
It turns out that such a pruning percentage limit is 10\% for both models in all the cases, except for conv 4 with 256 bits payload, where it is 20\%.

\begin{table*}[width=1.9\linewidth,cols=4,pos=h]
	\caption{Robustness against parameter pruning.}\label{pruning} 
	\begin{tabular}{c|c|c|ccc|ccc}
	\hline
	Embedded & Payload & $p$& \multicolumn{3}{c|}{W-DM-WRN}  & \multicolumn{3}{c}{W-WRN} \\
	\cline{4-9}
	layer & (bit) & & \multirow{2}{*}{TER (\%)}& TER  (\%)& BER (\%)& \multirow{2}{*}{TER (\%)}& TER  (\%)& BER (\%)\\
	&  &  & &after attack&  after attack& & after attack& after attack\\
	\cline{3-8}
	\hline
	Conv 2 & 256 &10\% & 8.20 & 10.20 & 0 & 8.15 & 10.23 & 0\\
\hline
	\multirow{2}{*}{Conv 3}& 256 &10\% & 8.15 & 8.83 & 0 & 8.02 & 8.49 & 0\\
	& 512 &10\% & 8.07 & 9.04 &0&8.30&8.80 &0 \\
\hline
	\multirow{2}{*}{Conv 4}& 256 & 20\% & 8.20 & 8.65 & 0 & 8.45 & 8.73 & 0\\
	& 4096 &10\%& 7.64 & 7.77 & 0 & 8.60 & 9.69 & 0\\
\hline
	multi-layer & 768 & 10\% &8.38 & 11.53 & 0 & 8.24 & 11.62 & 0\\
	\hline
	\end {tabular}
\end{table*}
From the table, we also observe that, under parameter pruning, a lower payload has to be considered for both the proposed method and Uchida et al's algorithm, especially in conv 2 and conv 3. With a larger payload, in fact, even if the BERs after pruning are still zero,  the TERs increase too much.

\begin{table}
	\caption{TER and BER of W-DM-WRNs and W-WRNs with different pruning percentages $p$.}
	\label{Different pruning percentage}
	\begin{tabular}{c|c c | c c}
		\hline
		\multirow{2}{*}{$p$} &  \multicolumn{2}{c|}{W-DM-WRNs} &  \multicolumn{2}{c}{W-WRNs}  \\ \cline{2-5}
		& TER(\%) & BER(\%) & TER(\%) & BER(\%)\\
		\hline
		10\% & 8.31 & 0 & 8.52 & 0\\
		20\% & 8.65 & 0 & 8.73 & 0\\
		30\%& 8.87 & 0 & 8.99 & 0\\
		40\%& 10.06 & 0 & 9.56 & 0 \\
		50\%& 12.37 & 0 & 10.86 &0\\
		60\%&19.23 & 0 & 15.35 &0\\
		70\%&25.95 &6.64&21.15&0\\
		\hline
	\end{tabular}
\end{table}
\section{Conclusions and final remark}
\label{sec.con}

In this paper, we proposed a new DNN watermarking algorithm that leverages on the watermarking with side information paradigm to decrease the obtrusiveness of the watermark and increase the payload. Inspired by the watermarking technique in \cite{Uchida17}, and  exploiting the ST-DM watermarking paradigm, we have proposed to use a new regularization term into the loss function for watermark embedding. Based on the experimental results we carried out, it turns out that the proposed method can reach a higher payload with a lower test error rate. We also verified that
the improvement is achieved without impairing the robustness of the models against fine tuning and parameter pruning.

An interesting direction for future research regards the security of the watermarking algorithm, that is the capability of the watermark to resist deliberate attempts to remove it.
For instance, assuming that the attacker is aware of the watermarking methodology, he may attempt to erase the original watermark by embedding a new one. Experiments could be carried out to assess the robustness against watermark overwriting both in the more favorable scenario where the attacker knows the position of the watermark and in the more realistic one where the attacker does not have any knowledge about the watermarked layers.

\appendix
\section{Acknowledgments}

This work has been partially supported by the China Scholarship Council(CSC), file No. 201907000056. The author would also thank Prof. Hongxia Wang (hxwang@ scu.edu.cn) for her valuable advice and help.

\printcredits



\bibliographystyle{cas-model2-names}

\bibliography{WatDNN}





\end{document}